\documentclass[amsmath,amssymb,aps,prl,reprint,superscriptaddress]{revtex4-1}

\usepackage{amsmath,amssymb}
\usepackage[english]{babel}
\usepackage{ bbold }
\usepackage{upgreek}
\usepackage{dsfont}
\usepackage{graphicx}
\usepackage{comment}
\usepackage{braket} 
\usepackage{colortbl}

\newcommand{\nsection}[1]{\vspace{0.2cm} \noindent{\bf #1}\newline}
\newcommand{\nsubsection}[1]{\vspace{0.1cm} \noindent{\bf #1}\newline}
\newcommand{\bls}[1]{\mathbf{#1}} %
\newcommand{\blg}[1]{\text{\boldmath$#1$}} %

\newcommand{\tcb}[1]{{#1}}

\begin{document}

\title{Quantum electromechanics with levitated nanoparticles}

\author{Lukas Martinetz}
\author{Klaus Hornberger}
\affiliation{University of Duisburg-Essen, Faculty of Physics, Lotharstra\ss e 1, 47048 Duisburg, Germany}
 \author{James Millen}
 \affiliation{King's College London, Department of Physics, Strand, London WC2R2LS, United Kingdom}
 \author{M. S. Kim}
 \affiliation{Imperial College London, Quantum Optics and Laser Science, Exhibition Road, London SW72AZ, United Kingdom}
\author{Benjamin A. Stickler}
\affiliation{Imperial College London, Quantum Optics and Laser Science, Exhibition Road, London SW72AZ, United Kingdom}
 \affiliation{University of Duisburg-Essen, Faculty of Physics, Lotharstra\ss e 1, 47048 Duisburg, Germany}

\begin{abstract}
\noindent
{\bf Preparing and observing quantum states of nanoscale particles is a challenging task with great relevance for quantum technologies and tests of fundamental physics. 
In contrast to atomic systems with discrete transitions, nanoparticles exhibit a practically continuous absorption spectrum 	and thus their quantum dynamics cannot be easily manipulated.
Here, we demonstrate that charged nanoscale dielectrics can be artificially endowed with a discrete level structure by coherently interfacing their rotational and translational motion with a superconducting qubit. We propose a pulsed scheme for the generation and read-out of motional quantum superpositions and entanglement between several levitated nanoparticles,
providing an all-electric platform for networked hybrid quantum devices.}
\end{abstract}

\maketitle

\nsection{Introduction}
Opto- and electromechanical systems are at the cutting edge of modern quantum devices \cite{rossi2018measurement,ockeloen2018stabilized,riedinger2018remote},  with great potential for technological application and fundamental  tests \cite{pikovski2012probing,aspelmeyer2014,khosla2018displacemon}.
Optically levitating nanoscale objects almost perfectly isolates them from their surroundings, enabling superior force sensitivity and coherence times \cite{millen2019optomechanics}.
Levitated nanoparticles have been successfully cooled into their motional quantum groundstate \cite{delic2019motional},
opening the door to free-fall quantum experiments \cite{romeroisart2011a,bateman2014,stickler2018probing}.

Quantum experiments with trapped nanoparticles require schemes to coherently control their rotational and translational quantum states. Continuous-wave optical techniques
are limited by the detrimental impact of photon scattering decoherence \cite{romeroisart2011b}
and by internal heating due to photon absorption \cite{millen2014nanoscale,hebestreit2018measuring}. In addition, the fact that nanoscale particles lack the discrete internal spectrum of atoms or other microscopic quantum systems makes it difficult to address them coherently with laser pulses. 

\begin{figure}
\centering
\includegraphics[width = 0.49\textwidth]{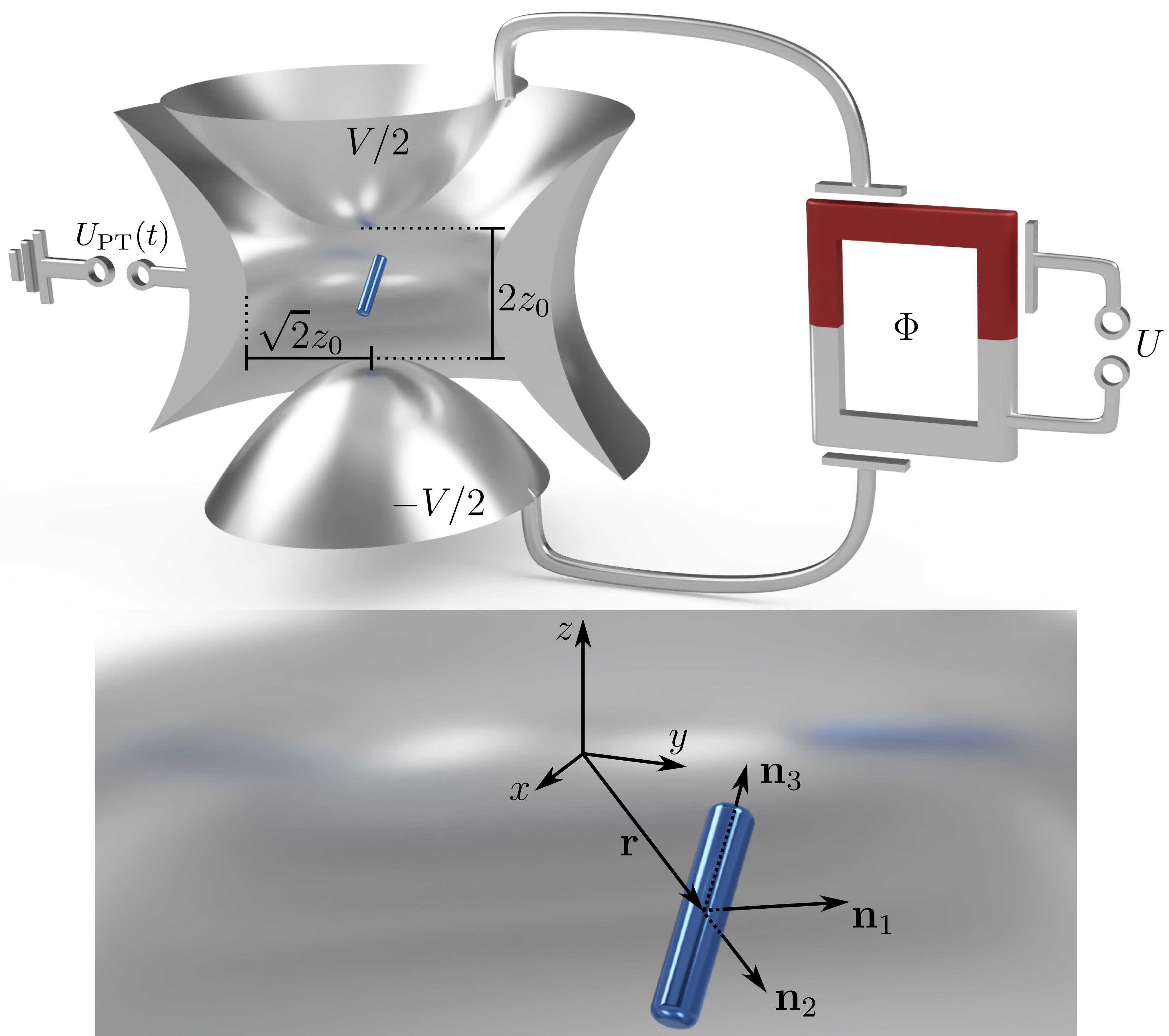}
\caption{The rotational and translational motion of a charged nanoparticle (blue) levitated in a Paul trap induces an electrical current between the superconducting endcap electrodes. The latter can be coherently interfaced with a charge qubit formed by a superconducting island (red). This Cooper-pair box can be used to generate and read-out spatial superpositions of the nanoparticle. The system constitutes the basic element for networking levitated nanoparticles into hybrid quantum devices based on superconducting circuitry.} \label{figure1}
\end{figure}

Here, we demonstrate that such a beneficial discrete level structure can be artificially introduced by coherently interfacing a charged nanoparticle levitated in a Paul trap with a superconducting qubit, through which its 
quantum dynamics can be manipulated and read out (see Fig.\,\ref{figure1}).  \tcb{In this all-electrical setup, the nanoparticle rotations decouple from the center-of-mass dynamics under experimentally realistic conditions, rendering 
it ideally suited for superposition experiments with a wide variety of particle geometries and charge distributions.}
\tcb {This paves} the way for networking nanoscale objects with superconducting quantum technologies. 

Quadrupole ion traps provide exceptionally stable confinement for  charged nanoparticles \cite{millen2015cavity}.
 \tcb{Moreover}, the particle motion  induces an electric current in the endcap electrodes.
We  propose to use this current for \tcb{first cooling the nanoparticle to milliKelvin temperatures and then} interfacing 
\tcb{its} motion
with a superconducting circuit. 
The resulting coupling between superconductor and particle \tcb{scales} as charge over root mass \cite{tian2004,goldwater2018levitated} \tcb{and can thus be as strong as for a single atomic ion for realistic charge distributions.}

We show that the proposed all-electrical platform is ideally suited for \tcb{nanoparticle cooling and interference experiments}
and for generating and reading-out  entanglement between several particles and superconducting qubits, thus forming a building block of a larger quantum network.
\tcb{The motional quantum state can be prepared and observed by qubit manipulations with}
an \tcb{ultra-fast} pulse scheme, \tcb{operating}
on a timescale much \tcb{shorter} than the mechanical period and within the coherence time of the charge qubit. \tcb{This pulse scheme allows to speed-up the observation of nanoscale quantum interference in a variety of opto- and electromechanical setups \cite{armour2002entanglement,scala2013matter,yin2013large,delord2018}.}

\nsection{Ro-translational macromotion}
\tcb{A} charged nanoparticle \tcb{is} suspended in a hyperbolic Paul trap of endcap distance $2 z_0$ and radius $\sqrt{2}z_0$, where the ring electrode is put to the time-dependent potential $U_{\rm PT}(t)=U_{\rm dc}  + U_{\rm ac}\cos (\Omega_{\rm ac}t)$ with respect to the floating endcaps, see Fig.\,\ref{figure1}.
Due to the quadrupole symmetry of the electric field, the rotational and translational particle motion is fully determined by its total charge $q$, orientation-dependent dipole vector ${\blg p}(\Omega)$, and quadrupole tensor ${\sf Q}(\Omega)$. Here, $\Omega$ denotes the orientational degrees of freedom of the particle, e.g.\ parametrized  by Euler angles; its center-of-mass position is ${\bf r}$.

\tcb{In general, the resulting time-dependent force and torque will lead to complicated and unstable dynamics of the nanoparticle. However, if}
the trap is driven sufficiently fast, \tcb{its} micro-motion can be separated, and \tcb{one obtains a time-independent} effective \tcb{trapping} potential for the macro-motion (see Methods)
\begin{eqnarray}
V_{\rm eff}({\bf r}, \Omega) & =&\frac{U_{\rm dc}}{2 z_0^2}\left(\frac{q}{2}\bls{r}\cdot\mathsf{A}\bls{r}+\blg{p}\cdot\mathsf{A}\bls{r}-\frac{1}{2}\bls{e}_z\cdot\mathsf{Q}\bls{e}_z\right) \notag \\
& &+\frac{U_{\rm ac}^2}{16 z_0^4\Omega_{\rm ac}^2} \sum_{i = 1}^3 \frac{1}{I_i} \left [ {\bf n}_i \cdot \left(\blg{p}\times\mathsf{A}\bls{r}+\bls{e}_z\times\mathsf{Q}\bls{e}_z\right) \right ]^2  \notag \\
& &+\frac{U_{\rm ac}^2}{16 M z_0^4\Omega_{\rm ac}^2}\left(q\bls{r}+\blg{p}\right)\cdot\mathsf{A}^2\left(q\bls{r}+\blg{p}\right).\label{eq:veff}
\end{eqnarray}
Here, $M$ is the particle mass and the Paul trap symmetry axis is aligned with ${\bf e}_z$, such that $\mathsf{A}=\mathbb{1}-3\bls{e}_z\otimes\bls{e}_z$. The $I_i$ denote the moments of inertia with ${\bf n}_i$ the associated directions of the rotor principal axes. We dropped the orientation dependence of ${\blg p}$, ${\sf Q}$, and ${\bf n}_i$ for compactness.

The effective potential \eqref{eq:veff} describes the coupled rotational and translational macromotion of an arbitrarily charged and shaped nanoparticle in a quadrupole ion trap, \tcb{and is thus pertinent for  ongoing nanoparticle experiments \cite{millen2015cavity,delord2018}}. \tcb{It} shows that stable trapping can be achieved for sufficiently small bias voltages $U_{\rm dc}$ (\tcb{with frequencies $\omega_z = q U_{\rm ac}/ \sqrt{2} M \Omega_{\rm ac}z_0^2$ and $\omega_{x,y} = \omega_z/2$}). \tcb{In addition,} the rotational and translational motion decouple for particles with vanishing dipole moment. Note that if the particle has a finite quadrupole moment its  rotation dynamics can still be strongly affected by the trapping field.

\nsection{Particle-circuit coupling}
The rotational and translational motion of the particle induces mirror charges in the endcap electrodes. \tcb{The latter can be quantified by}
extending the Shockley-Ramo theorem to arbitrary charge distributions (see Methods), \tcb{yielding the capacitor charge} $Q=-k\bls{e}_z\cdot(q\bls{r}+\blg{p})/z_0+CV$, given the endcap capacitance $C$ and voltage drop $V$. The geometry factor $k$, with values of $0 < k \lesssim 1/2$ for realistic electrode geometries,   determines the approximately homogeneous field $-k V/z_0 \,{\bls e}_z$ close to the trap center in absence of the particle. (A perfect plate capacitor corresponds to $k=1/2$.) 

The induced capacitor charge only depends on the total motional dipole moment $q \bls{r} + \blg{p}$ along the Paul trap axis, \tcb{and is thus independent of the particle quadrupole moment}. A circuit connecting the electrodes picks up the ro-translational motion of the particle via the current $I = dQ/dt$. At the same time, the particle feels a voltage-dependent electrostatic force and torque depending on the circuit state. This can be used \tcb{for resistive cooling and for coherently interfacing} the particle with a superconducting qubit.
\tcb{Nanoparticle resistive cooling} can be \tcb{achieved} by joining the endcaps with a resistance $R$. %
The dissipation of the induced current in the resistor leads to thermalization of the particle motion at the circuit temperature. The timescale of this resistive cooling can be tuned by adding an inductance in series to the circuit \cite{goldwater2018levitated}. In the adiabatic limit, it reacts almost instantaneously to the particle motion. Thus, the circuit degrees of freedom can be expanded to first order in the particle velocity and rotation speed, yielding the effective total cooling rate
\begin{equation}
\gamma_{\rm ad}=\frac{Rk^2}{z_0^2}\left(\frac{q^2}{M}+ \sum_{i = 1}^3 \frac{1}{I_i} \left [ {\bf n}_i \cdot (\blg{p}\times\bls{e}_z) \right ]^2\right). \label{resistivecoolingrate}
\end{equation}
This quantifies how fast an initially occupied phase-space volume contracts, predicting the timescale of rotational and translational thermalization with the circuit.
The rate \eqref{resistivecoolingrate} is always positive and exhibits a $q^2/M$-scaling, 
\tcb{indicating that charged nanoparticles can be cooled as efficiently as atomic ions.}

\nsection{Interfacing nanoparticle and charge qubit}
The levitated nanoparticle \tcb{can be coherently coupled to} a superconducting Cooper-pair box \tcb{by attaching the latter to the endcap electrodes (see Fig. \ref{figure1}).} \tcb{The nanoparticle motion towards the endcaps modifies the voltage drop over the Cooper-pair box, whose charge state determines the force and torque acting on the particle. Preparing the Cooper-pair box in a superposition of charge states thus entangles the nanoparticle motion with the circuit. This can be used to generate and verify nanoscale motional superposition states.}

\tcb{The combined nanoparticle-Cooper-pair box Hamiltonian can be derived in a lengthy calculation from Kirchhoff's circuit laws (see Methods). Operating the Cooper-pair box in the charge qubit regime of $N$ and $N+1$ Cooper pairs yields the nanorotor-qubit coupling 
\begin{equation}
H_{\rm int} = - \frac{2 e k}{C_\Sigma z_0} (N + \sigma_+ \sigma_-  ) (q {\bf r} + \blg{p} )\cdot {\bf e}_z,
\end{equation}
where $C_\Sigma$ is the effective capacitance of the circuit and the qubit raising and lowering operators are denoted by $\sigma_+$, $\sigma_-$. This interaction couples the charge eigenstates of the box to the motional dipole moment of the nanorotor, implying that charge states are conserved by the interaction and that rotations of the quadrupole or higher multipole moments of the nanoparticle are not coupled by the qubit.}

\tcb{In the experimentally realistic situation that the nanoparticle is almost homogeneously charged and inversion symmetric, its dipole moment is negligibly small. }
The rotational and translational \tcb{macromotion in the Paul trap \eqref{eq:veff}} then decouple even for large quadrupole moments. The center-of-mass motion along the Paul trap axis further decouples from the transverse degrees of freedom, \tcb{since only the motion towards the electrode is affected by the Cooper-pair box}. The particle trapping potential in $z$-direction is slightly shifted and stiffened due to the charge qubit (with $N\neq 0$), yielding the effective Hamiltonian
\begin{equation}
H_{\rm 1D}=E_{\rm c}\sigma_+\sigma_-+\hbar\omega a^\dagger a -\hbar\kappa\sigma_+\sigma_-\left(a + a^\dagger\right),\label{hamiltonianwechselwirkung}
\end{equation}
with charge energy $E_{\rm c}$ and coupling strength $\kappa=2ekq/C_\Sigma z_0\sqrt{2M\hbar\omega}$, where the nanoparticle oscillation with frequency $\omega^2=\omega_z^2 + q^2 k^2/C_\Sigma M z_0^2$ is described by the ladder operators $a$, $a^\dagger$ (see Methods). 

The Hamiltonian \eqref{hamiltonianwechselwirkung} demonstrates that the qubit can be used to generate nanoparticle quantum states. The absence of discrete internal transitions can thus be compensated by the 
non-linearity provided by a superconducting circuit. The nanoparticle-qubit coupling strength is proportional to charge over root mass, yielding appreciable coupling for highly charged nanoscale objects.

\tcb{Note that} a finite bias voltage $U_{\rm dc}$ applied to the ring electrode will not affect the Cooper-pair box, but produce an additional, approximately linear potential at the \tcb{shifted trap center $z_s$}. It adds the term $-V_{\rm ext} (a+a^\dagger)$ to (\ref{hamiltonianwechselwirkung}), with $V_{\rm ext}=qU_{\rm dc} z_s \sqrt{\hbar/2M\omega z_0^4}$. \tcb{This term will be used below to control the relative phase of the nanoparticle superposition state.}

\begin{figure*}
	\centering
	\includegraphics[width = \textwidth]{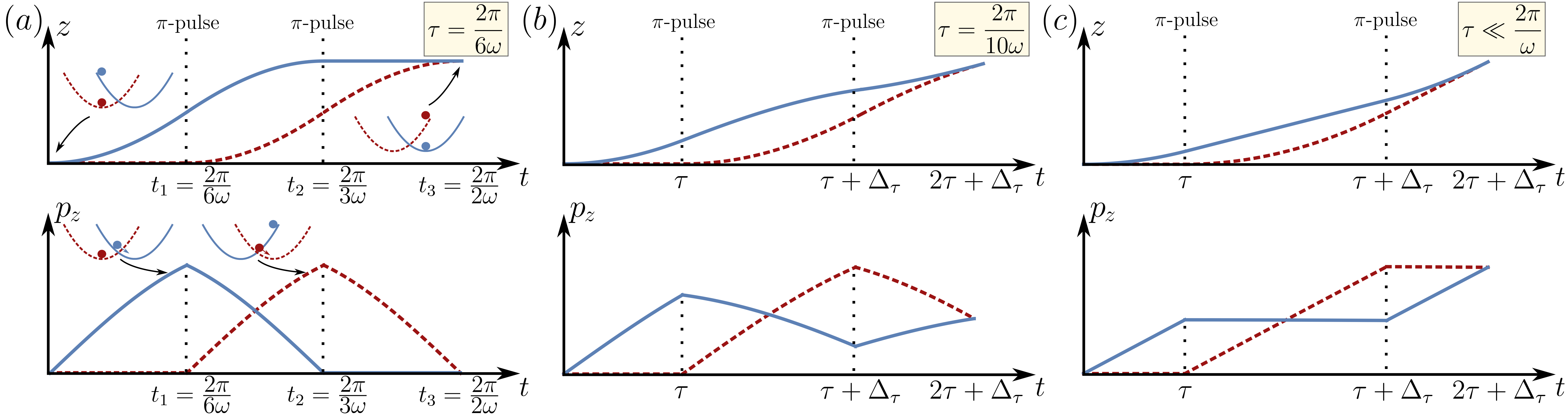}
	\caption{Position and momentum trajectories of the proposed interference scheme. An initial $\pi/2$-pulse on the Cooper-pair box generates a charge superposition in the endcap electrodes, so that the charged nanoparticle feels a superposition of a spatially shifted and an unshifted harmonic potential. The time evolution (\ref{hamiltonianwechselwirkung}) gives rise to two wave packets traveling on separate trajectories. 
		To verify this motional superposition state the wave packets must be reunited.
		This can be achieved by applying two $\pi$-pulses, each interchanging the potentials felt by the two branches, in such a way that the trajectories finally coincide in position and momentum.
		(a)
		In the simplest case all pulses are separated by one sixth of the harmonic oscillation period and the particle is initially at rest. The $\pi/2$-pulse then leaves one branch unaffected (red dashed line), while the other one is accelerated (blue line). The first $\pi$-pulse accelerates the resting branch and decelerates the moving one to a standstill at the time of the second $\pi$-pulse. After that, the blue trajectory remains at rest while the red one is decelerated until it reaches the blue one with zero velocity.
		(b) Even for arbitrary pulse times $\tau$ the 
		separation $\Delta_\tau$ between the $\pi$-pulses 
		can be chosen such that the corresponding paths in phase space coalesce for all initial states at $2\tau+\Delta_\tau$.
		(c) The scheme works for time durations much shorter than the oscillation period, which makes it particularly suitable for limited coherence times. In this short time limit the accelerations are essentially constant and $\Delta_\tau=2\tau$.
	} \label{figure2}
\end{figure*}

\nsection{Generating and observing superpositions}
Quantum interference of the nanoparticle motion \tcb{on short timescales} can now be performed by a \tcb{rapid} sequence of qubit rotations and measurements.
At the beginning of the interference scheme, the charge qubit is prepared in its groundstate $\ket{N}$, while the nanoparticle is cooled to temperature $T$, $\rho_0 = \ket{N}\bra{N}\otimes\exp ( - \hbar \omega a^\dagger a/k_{\rm B} T)/Z $. 
After this initial state preparation $V_{\rm ext}$ is switched to a constant value. The free dynamics governed by  (\ref{hamiltonianwechselwirkung}) with the external potential is then intersected by $\sigma_x$-rotations of the qubit at {\em four} different times:
\begin{itemize}
\setlength\itemsep{-0.2em}
\item[(i)]\tcb{a $\pi/2$-pulse at $t=0$, which prepares the qubit in a superposition of charge states},
\item[(ii)] \tcb{a $\pi$-pulse at $t=t_1$, which flips the qubit state,}
\item[(iii)] \tcb{another $\pi$-pulse at $t=t_2$, and}
\item[(iv)] \tcb{a $\pi/2$-pulse at $t=t_3$ with subsequent measurement of the qubit occupation ${\sigma_+\sigma_-}$.}
\end{itemize}

With a symmetric pulse scheme, i.e.\ $t_1 = t_3-t_2=\tau $, it is always possible to find a  $ \Delta_\tau\equiv t_2-t_1 $ such that the nanoparticle state evolves into a superposition and then recombines with maximal overlap (see Methods). The corresponding phase space trajectories are illustrated in Fig.~\ref{figure2}. 
In this case, \tcb{the particle motion is first entangled with the qubit, generating} a motional quantum superposition. \tcb{This superposition is then} reversed by steps (ii) and (iii), and finally recombined, \tcb{undoing the entanglement}. Through this sequence, the phase imprinted on the nanoparticle motion through the external voltage $V_{\rm ext}$ is transferred onto the qubit state and read-out via its population, 
\begin{eqnarray}\label{measurementoutcome2}
\braket{\mathsf{\sigma}^+\mathsf{\sigma}^-} &=&\cos^2\left[\left(\frac{\kappa^2}{\omega} + \frac{2 \kappa V_{\rm ext}}{\hbar \omega}-\frac{E_{\rm c}}{\hbar}\right) \left ( \tau - \frac{\Delta_\tau}{2}\right ) \right].
\end{eqnarray}
Varying $V_{\rm ext}$ and observing the corresponding modulation of the qubit population can thus be used to verify 
that the nanoparticle existed in a spatial superposition state. The final population also oscillates as a function of  the pulse time $\tau$.

\tcb{This pulse scheme enables the generation and observation of nanoscale quantum superpositions in harmonic potentials with} pulse separations $\tau$ 
much shorter than the particle oscillation period. 
It is therefore applicable to various opto- and electromechanical  systems \cite{armour2002entanglement,scala2013matter,delord2018}.
The  required accuracy of the pulse times,  determined by the qubit frequency and the particle temperature $T$, must ensure that the phase $\kappa V_{\rm ext} (2\tau-\Delta_\tau)/ \hbar \omega$ is measurable.

\begin{figure}
	\centering
	\includegraphics[width = 0.5\textwidth]{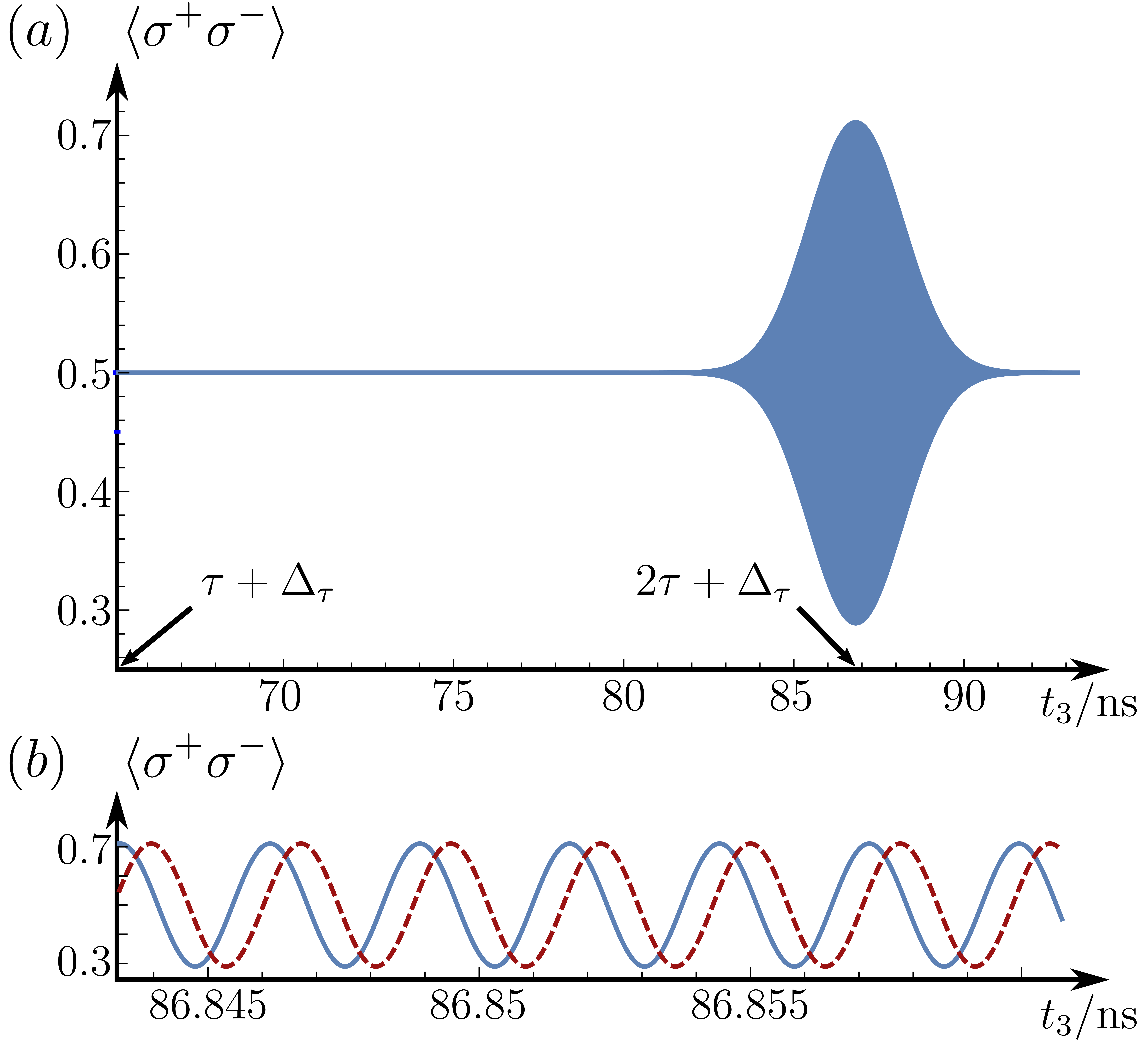}
	\caption{
		The Cooper-pair box occupation (\ref{measurementoutcome})
		shows a pronounced interference signal for experimentally realistic parameters if $\pi$-pulses are applied at $t_1=21.7\,$ns and $t_2=65.1\,$ns. The interfering trajectories then coalesce at $86.9\,$ns. (a) The envelope of the interference pattern is determined by the overlap of the two associated nanoparticle wave packets. Its width decreases with increasing temperature. Even at  1\,mK, corresponding a mean phonon number of 945, more than a thousand fringes can be expected (assuming a qubit dephasing time $1/\gamma_{\rm d}=100\,$ns, see methods).
		(b) The frequency of the interference signal is mainly determined by the charge energy $E_{\rm c}$ of the qubit. A bias voltage $U_{\rm dc}=5\,$V applied on the ring electrode imprints a phase on the nanoparticle, which shifts the interference pattern by about $2\pi/5$ (dashed line). 
	} \label{figure3}
\end{figure}

To illustrate that nano- to microsecond motional superpositions can be realistically prepared and observed on the coherence time scale of a charge qubit \cite{houck2009life}, we show in Fig.~\ref{figure3} the expected interference signal of a $10^6$ amu particle at $T=1$\,mK. The nanoparticle is assumed to be cylindrically shaped, with a homogeneous surface charge of $q=200$\,e and a realistic dipole moment (see Methods). It is stably levitated inside a sub-millimeter Paul trap, its motion well approximated by a harmonic oscillation with $\omega=138\,$kHz.
We find that the resulting strong coupling to the Cooper-pair box of $\kappa=16.8\,$MHz  renders the nanoparticle particle sensitive to the presence or absence of a single Cooper pair. A voltage of $U_{\rm dc}=25\,$V then suffices to imprint a relative phase on the motional superposition that shifts the interference pattern by a full fringe.

\nsection{Networking levitated nanoparticles}
The proposed interference protocol can be extended to transfer qubit entanglement to nanoparticles. Levitated objects may thus be coherently integrated into superconducting quantum networks, e.g.\ for sensing and metrology applications. Here we illustrate how to entangle the nanoparticle with a second, separated charge qubit, or with another nanoparticle levitating in a distant Paul trap.

Entanglement of the nanoparticle with a second qubit is achieved by replacing the initial $\pi/2$-pulse with an operation which prepares a maximally entangled two-qubit state \cite{yamamoto2003demonstration, rodrigues2008entanglement}.
To verify the involvement of the particle in the nonlocal dynamics one carries out the above interference protocol by performing all pulses on both qubits.
The occupation of the separated qubit, conditioned on having found the directly coupled one in the groundstate, is then given by
\begin{eqnarray}
\braket{\sigma^+\sigma^-} &=&\cos^2\left[\chi  \left (\tau - \frac{\Delta_\tau}{2} \right )\right],  \label{10}
\end{eqnarray}
with $\chi=\kappa^2/\omega + 2 \kappa V_{\rm ext}     /\hbar \omega -(E_{{\rm c1}}-E_{{\rm c2}})/\hbar$. This assumes that the qubits were initially prepared in the 
singlet state $\ket{\Psi^-}$. %
The external potential $V_{\rm ext}$,  acting only on the nanoparticle, thus serves to fully control the measurement outcome of the distant qubit.
Having established that the coherent dynamics extends from the particle to the distant qubit, entangled states
of these two systems can be produced by carrying out step (iv) and the subsequent measurement of the directly coupled qubit at $t_3<2\tau+\Delta_\tau$, i.e.\ before the particle wave packets overlap.

An all-electrical protocol to entangle two distant levitated nanoparticles works along the same lines:
We consider two distant nanoparticle-qubit setups of identical frequency $\omega$, where the qubits are again initially in the state $\ket{\Psi^-}$.
To verify the involvement of both particles in the nonlocal dynamics,
one carries out the protocol until the time $t_3=2\tau+\Delta_\tau$ of wave packet overlap. The occupation  of the second qubit, conditioned on having found the first one in the ground state, is then given by (\ref{10}), with $\chi=\chi_1-\chi_2$, where $\chi_i = \kappa_i^2/\omega + 2 \kappa_i V_i /\hbar \omega -E_{{\rm c}i}/\hbar$. The interference pattern thus depends on the difference of the local nanoparticle phases. 
By measuring both qubits before the wave packets overlap, e.g. at $\tau+\Delta_\tau<t_3<2\tau+\Delta_\tau$ the two oscillators can be projected onto an entangled motional state (see Methods).

\nsection{Conclusion}
The coherent control of charged nanoparticles by superconducting qubits offers a new avenue for quantum superposition experiments with massive objects. \tcb{The nanoparticle superposition state is generated and read-out by pulsed qubit rotations and measurements, enabling interference experiments on ultra-short time scales. All-electric trapping, cooling, and manipulation}
avoids photon scattering and absorption, the dominant decoherence sources in laser fields. \tcb{In addition, the nanoparticle rotations decouple from the centre-of-mass motion for realistic particle shapes and charge distributions, rendering this setup widely applicable. It} holds the potential of bridging the mass gap in quantum superposition tests from 
current experiments with 
massive molecules \cite{fein2019quantum} to future guided interferometers with superconducting microscale particles \cite{pino2018chip}.
Moreover, these novel hybrid quantum devices can serve as building blocks for larger networks connected by superconducting circuitry, distributing entanglement between multiple nanoparticles. 

The presented qubit-nanoparticle coupling scheme is feasible with available technology for realistic particles. 
Beyond that, the degree of quantum control can be enhanced by fabricating particles with tailored dipole and quadrupole moments and by combining electric with optical techniques \cite{tebbenjohanns2019cold,conangla2019optimal}. 
This may give rise to the observation of coherent effects between the rotational and translational nanoparticle degrees of freedom, and provide a platform for studying charge-induced decoherence in an unprecedented mass and complexity regime.

\bibliographystyle{apsrev4-1}
%

\vspace*{\baselineskip}
{\em\noindent
	Acknowledgements:}
LM, KH, and BAS acknowledge funding from the Deutsche Forschungsgemeinschaft (DFG, German Research Foundation) -- 411042854. BAS acknowledges funding from the European Union's Horizon 2020 research and innovation programme under the Marie Sklodovska-Curie grant agreement No 841040. 
MSK thanks the Royal Society, the UK EPSRC (EP/R044082/1) and KIST Open Research Program. JM is supported by the European Research Council (ERC) under the European Union's Horizon 2020 research and innovation programme (Grant agreement No. 803277), and by EPSRC New Investigator Award EP/S004777/1. 
\vspace*{\baselineskip}

\vspace*{\baselineskip}
{\em\noindent Author contributions:}
All authors contributed conceptually to the proposal. LM, KH, and BAS performed the analytic calculations and wrote the manuscript with input from JM and MSK. 

\vspace*{\baselineskip}
{\em\noindent Data availability:}
No data sets were generated or analysed during the current study.

\vspace*{\baselineskip}
{\em\noindent Competing interests:}
The authors declare no competing interests.

\vspace*{\baselineskip}
\noindent
{\bf Methods}

\small

\nsubsection{Ro-translational macromotion in a Paul trap}
An arbitrarily charged nanoparticle moving and revolving at position ${\bls R}$ and orientation $\Omega$ in a hyperbolic Paul trap is subject to the time-dependent potential
\begin{equation}
V({\bls R},\Omega,t) =\frac{U_{\rm PT}(t)}{2z_0^2}\left(\frac{q}{2}\bls{R}\cdot\mathsf{A}\bls{R}+\blg{p}\cdot\mathsf{A}\bls{R}-\frac{1}{2}\bls{e}_z\cdot\mathsf{Q}\bls{e}_z\right),\label{VPaultimedependent}
\end{equation}
with $\mathsf{A}=\mathbb{1}-3\bls{e}_z\otimes\bls{e}_z$. Here, the dipole moment $\blg{p}$ and the quadrupole tensor $\mathsf{Q}$ depend on the principal axes ${\bf N}_i$ of the nanoparticle with the associated moments of inertia $I_i$. The effective potential for the macromotion is obtained by setting ${\bls R} = {\bf r} + \blg{\epsilon}$, 
and ${\bf N}_i = {\bf n}_i + \blg{\delta} \times {\bf n}_i$, serving to separate the center-of-mass macromotion ${\bf r}$ from the much faster micromotion $|{\blg \epsilon}| \ll |{\bf r}|$ varying with zero mean. Similarly, the rotational micromotion  $|{\blg \delta}| \ll 1$ varies much faster than ${\bf n}_i$.
The center-of-mass and angular momentum obey
\begin{subequations} \label{15}
\begin{equation}
m\ddot{\bls{R}}=-\frac{U_{\rm PT}(t)}{2z_0^2}\mathsf{A}\left(q\bls{R}+{\blg p}\right) \label{appendixexactforce}
\end{equation}
and
\begin{eqnarray}
\dot {\bf J} =-\frac{U_{\rm PT}(t)}{2z_0^2}\left(\blg{p}\times\mathsf{A}\bls{R}+\bls{e}_z\times{\sf Q} \bls{e}_z\right).\label{appendixexacttorque}
\end{eqnarray}
\end{subequations}
Taking macromotion to be approximately constant on the time scale of the micromotion and neglecting all small terms yields 
\begin{subequations} \label{16}
\begin{equation}
\blg{\epsilon}\approx\frac{U_{\rm ac}\cos(\Omega_{\rm ac} t)}{2M z_0^2 \Omega_{\rm ac}^2}\mathsf{A}\left(q\bls{r}+\blg{p}\right)
\end{equation}
and
\begin{equation}
\blg{\delta}\approx\frac{U_{\rm ac}\cos(\Omega_{\rm ac} t)}{2 z_0^2 \Omega_{\rm ac}^2}\sum_{i = 1}^3 \frac{1}{I_i} {\bf n}_i [{\bf n}_i \cdot \left(\blg{p}\times\mathsf{A}\bls{r}+\bls{e}_z\times\mathsf{Q}\bls{e}_z\right)],
\end{equation}
\end{subequations}
involving the familiar Mathieu parameter and its rotational analogues, respectively. The dipole and quadrupole moments here only include the macromotion, i.e.\ $\bls{p}=\sum p_i \bls{n}_i$ and $\mathsf{Q}=\sum Q_{ij}\bls{n}_i\otimes\bls{n}_j$, in contrast to (\ref{15}).  

The effective force and torque of the macromotion can be obtained by inserting (\ref{16}) into (\ref{15})  and averaging over one micromotion cycle. A lengthy but straightforward calculation demonstrates that they can be expressed through the time-independent effective potential \eqref{eq:veff}. 
We remark that the same potential \eqref{eq:veff} can also be derived quantum mechanically by adapting the method outlined in Ref.~\cite{cook1985quantum} for the combined rotational and translational motion of the nanoparticle.

\nsubsection{Generalized Shockley-Ramo theorem}
\noindent
To calculate the current induced by an arbitrary, rigidly bound charge distribution moving and rotating between the endcap electrodes we use Green's reciprocity theorem, 
\begin{equation}
\int_{\mathcal{V}}dV \phi_{\rm ref}\rho+\int_{\partial \mathcal{V}}dA \phi_{\rm ref}\sigma=\int_{\mathcal{V}}dV \phi\rho_{\rm ref}+\int_{\partial \mathcal{V}}dA \phi\sigma_{\rm ref}.
\end{equation}
It relates the particle charge density $\rho$, the electrode surface charge density $\sigma$ and the electrostatic potential $\phi$ to those of a reference system. 
Choosing the reference system to have no particle in the trap volume $\mathcal{V}$, a vanishing potential on the ring electrode, and opposite potentials on the endcaps leads to an approximately linear potential $\phi_{\rm ref}$ near the trap center. This results in
\begin{equation} \label{eq:shra}
\frac{k}{z_0}\bls{e}_{z}\cdot\int_\mathcal{V} dV \bls{x}\rho(\bls{x})+\frac{Q_1-Q_2}{2}=CV,
\end{equation}
where $Q_1$ and $Q_2$ are the total charges on the top and bottom endcap and $\bls{x}$ originates from the trap center. The remaining integration yields the \tcb{capacitance charge $Q = (Q_1 - Q_2)/2=-k\bls{e}_z\cdot(q\bls{r}+\blg{p})/z_0+CV$}, and its time derivative the induced current.

\begin{figure*}
\centering
\includegraphics[width = 0.75\textwidth]{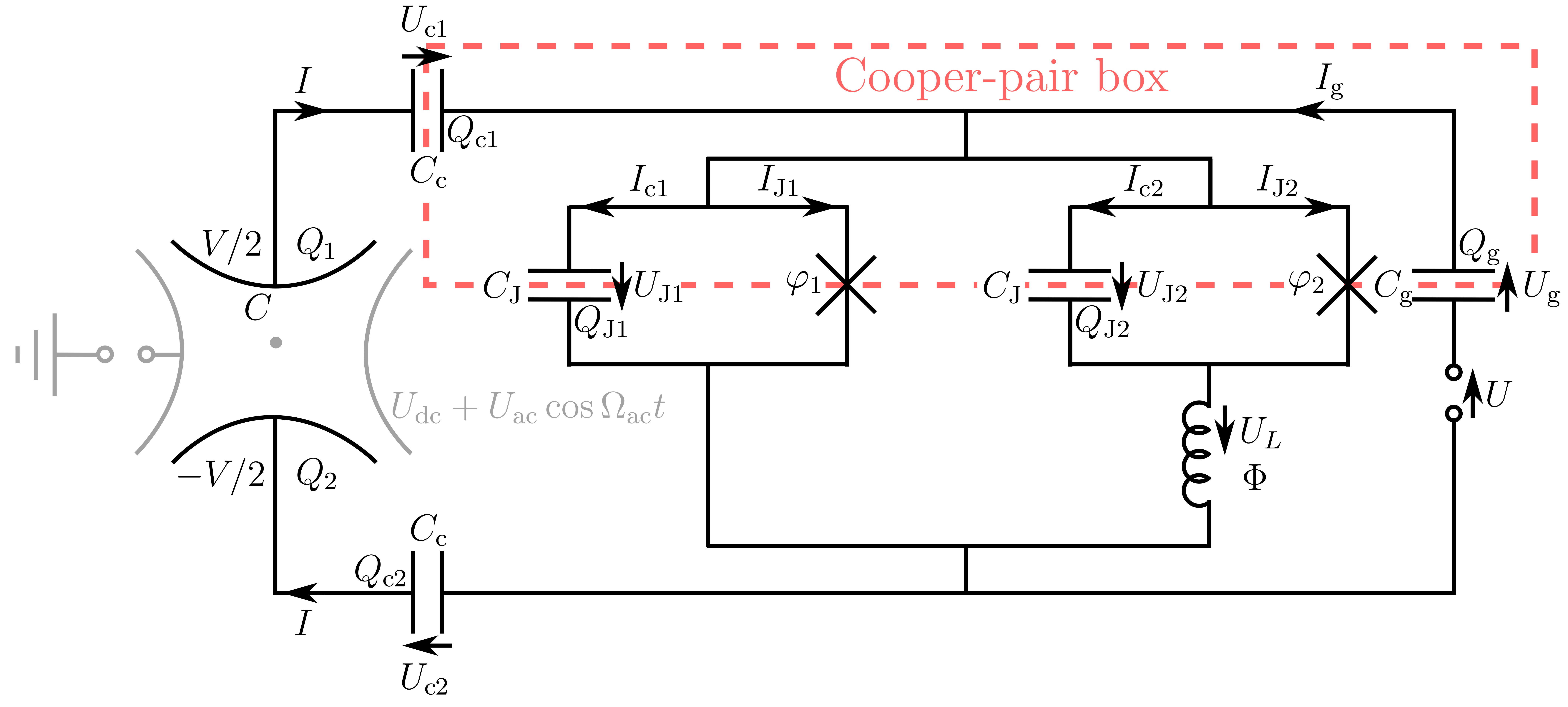}
\caption{Circuit diagram of the tunable Cooper-pair box coupled to the particle via the trap capacitor. 
	The superconducting island to and from which Cooper pairs can tunnel is indicated by the dashed line.
	The arrows 
	indicate the direction of increasing electrostatic potential for positive $U$'s and of the electron flow   for positive $I$'s.
} \label{AppendixFig1}
 
\end{figure*}

\nsubsection{Cooper-pair box-nanoparticle Hamiltonian}
\noindent
Figure~\ref{AppendixFig1} shows how the Paul trap is connected with the circuit. The latter consists of a superconducting loop with two Josephson junctions modeled as a capacitance $C_{\rm J}$ and a tunneling junction in parallel. The loop of vanishing inductance  encloses an external magnetic flux $\Phi$.

The quantum state of the circuit is described by a macroscopic wave function whose phase jumps $\varphi_1$, $\varphi_2$  at the Josephson junctions satisfy
$\varphi_2-\varphi_1+{2\pi\Phi}/{\hbar}=2\pi m$
with  $m$ an integer. Josephson's equations $I_{{\rm J}i}=I_{\rm c}\sin\varphi_i$ and 
$U_{{\rm J}i}=\frac{\hbar}{2e}\dot{\varphi_i}$ relate them to the tunneling current and the voltage drop in each junction $i = 1,2$. The loop is coupled capacitatively  to the  endcaps via $C_{\rm c}$, and can be controlled with \tcb{the external voltage $U$, applied via the gate capacitance $C_{\rm g}$}.
The circuit equations of motion can be obtained starting from 
Kirchhoff's laws,
\begin{subequations}
\begin{eqnarray}
V+U_{{\rm J1}}+U_{\rm c1}+U_{\rm c2} & = & 0,\label{kirchhoffa}\\
U_{\rm J1}+U+U_{\rm g} & = & 0, \label{kirchhoffb}\\
I_{\rm c1}+I_{\rm J1}+I_{\rm c2}+I_{\rm J2} & = & I+I_{\rm g}. \label{kirchhoffc}
\end{eqnarray}
\end{subequations}
Inserting \tcb{the capacitance charge \eqref{eq:shra}} into \eqref{kirchhoffa}, differentiating with respect to time and using that $U_{{\rm c}i}=Q_{{\rm c}i}/C_{\rm c}$ and that $I=\dot{Q}=\dot{Q}_{{\rm c}i}$ yields
\begin{equation}
\dot{Q}=-\frac{k_0}{z_0}\frac{C_{\rm eff}}{C}\left(q\dot{{\bf r}}+\dot{\blg{p}}\right)\cdot{\bf e}_z-\frac{C_{\rm eff}}{C_{\rm J}}\dot{Q}_{\rm J1},\label{appendixQdot}
\end{equation}
with the effective capacitance $C_{\rm eff}=CC_{\rm c}/(C_{\rm c}+2C)$. In addition, \eqref{kirchhoffb} and \eqref{kirchhoffc} yield the  relations
\begin{eqnarray}
\dot{Q}_{\rm g} & = & -\frac{C_{\rm g}}{C_{\rm J}}\dot{Q}_{\rm J1}-C_{\rm g}\dot{U}, \label{appendixQGdot} \\
\dot{Q}+\dot{Q}_{\rm g} & = & \dot{Q}_{\rm J1}+\dot{Q}_{\rm J2}+I_{\rm c}\sin{\varphi_1}+I_{\rm c}\sin{\varphi_2}.\label{appendixcurrentequation}
\end{eqnarray}
Inserting \eqref{appendixQdot} and \eqref{appendixQGdot} into \eqref{appendixcurrentequation}, using flux quantization, and defining $\varphi=\varphi_1-e\Phi/\hbar$ finally yields the circuit equation of motion,
\begin{equation}
\begin{split}\label{appendixEquationmotionphi}
\ddot{\varphi}=&-\frac{4eI_{\rm c}}{\hbar C_\Sigma}\cos\left(\frac{e\Phi}{\hbar}\right)\sin\varphi-\frac{2 e C_{\rm eff}k}{\hbar CC_\Sigma z_0}\left(q\dot{{\bf r}}+\dot{\blg{p}}\right)\cdot {\bf e}_z\\
&-\frac{2e}{\hbar C_\Sigma}\left[C_{\rm g}\dot{U}+\left(C_{\rm eff}+C_{\rm g}\right)\frac{\ddot{\Phi}}{2}\right],
\end{split}
\end{equation}
with  $C_\Sigma=C_{\rm eff}+C_{\rm g}+2C_{\rm J}$.

The ro-translational motion of the particle is driven by the endcap voltage $V$ via the force ${\bls F}=-kVq\bls{e}_z/z_0$ and the torque ${\bls N}=kV\bls{e}_z \times\blg{p}/z_0$, in addition to the Paul trap force and torque. In the relevant limit of large $C_{\rm c}$ the Hamiltonian generating the coupled dynamics of circuit and particle takes the form
\begin{eqnarray} \label{eq:hamtot}
H &=&\frac{2e^2}{C_\Sigma}\left [\frac{\Pi}{\hbar}-\frac{k}{2ez_0}(q {\bf r}+{\blg p})\cdot {\bf e}_z-n_{\rm g}\right ]^2-E_J\cos\varphi \notag \\
&&-\frac{k \dot\Phi}{2z_0}\left(q {\bf r}+{\blg p}\right) \cdot {\bf e}_z+H_{\rm rb}+V_{\rm eff}(\bls{r},\Omega)\,,
\end{eqnarray}
where the canonical momentum $\Pi$ conjugate to $\varphi$ quantifies the number of Cooper pairs on the island and $H_{\rm rb}$ is the free rigid body Hamiltonian for the center-of-mass motion and rotation. Eq.~(\ref{eq:hamtot})
involves the voltage-induced number of Cooper pairs $n_{\rm g}=C_{\rm g}U/2e+\left(C+C_{\rm g}\right)\dot\Phi/4e$, and the Josephson energy $E_{\rm J}=\hbar I_{\rm c}\cos\left(e\Phi/\hbar\right)/e$.
We choose the flux $\Phi$ and applied voltage $U$ so that the Cooper-pair box can be treated as an effective  two-level system \cite{nakamura1999coherent} with $N$ or $N+1$ Cooper pairs on the island, 
$\Pi = \hbar (N + \sigma_+\sigma_-)$. Tuning $n_{\rm g}$, $E_{\rm J}$, and $\dot\Phi$ to zero yields the Hamiltonian
\begin{eqnarray} \label{eq:qurot}
H & = & \frac{2 e^2}{C_\Sigma} (2 N + 1) \sigma_+ \sigma_- - \frac{2 e k}{C_\Sigma z_0} (N + \sigma_+ \sigma_-  ) (q {\bf r} + \blg{p} )\cdot {\bf e}_z \notag \\
&& + \frac{k^2}{2 C_\Sigma z_0^2} [(q {\bf r} + {\blg p})\cdot {\bf e}_z]^2 + H_{\rm rb} + V_{\rm eff}({\bf r},\Omega),
\end{eqnarray}
which leaves the  charge eigenstates of the box unaffected. The charge-dependent potential shift given by the second term will drive the particle into a ro-translational superposition if the charge states are superposed. 

Neglecting the dipole moment and separating the nanoparticle transverse motion and rotations finally yields \eqref{hamiltonianwechselwirkung} with the potential minimum shifted to $z_{\rm s}=2ek Nq/C_\Sigma z_0 M\omega^2$ and the charge energy $E_c=2e^2(1+2N-kqz_{\rm s}/ez_0)/C_\Sigma$.

\nsubsection{Time evolution and measurement outcome}
\noindent
The time evolution generated by  \eqref{hamiltonianwechselwirkung} \tcb{with external potential $V_{\rm ext}$} can be written, up to a global phase, as a combination of a qubit-dependent phase, qubit-dependent particle displacements and the free time evolution of the harmonic oscillator,
\begin{eqnarray} \label{eq:un}
U(t)& = &\exp\left[-it\left(\frac{E_{\rm c}}{\hbar}-\frac{\kappa^2}{\omega} - \frac{2 \kappa V_{\rm ext}}{\hbar \omega} \right)\sigma_+\sigma_-\right]\notag \\
&&\times\exp\left[\left (\frac{\kappa}{\omega} \sigma_+\sigma_-+\frac{V_{\rm ext}}{\hbar \omega}\right )\left(a^\dagger - a\right)\right] \notag \\
&& \times \exp\left(-i\omega t a^\dagger a \right) \notag \\
&&\times \exp\left[-\left (\frac{\kappa}{\omega} \sigma_+\sigma_-+\frac{V_{\rm ext}}{\hbar \omega}\right )\left(a^\dagger - a\right)\right].
\end{eqnarray}
The qubit is initially prepared in its groundstate while the nanoparticle is in a thermal state of temperature $T$. A $\pi/2$-pulse rotates the qubit into the superposition $\left(\ket{N}+i\ket{N+1}\right)/\sqrt{2}$, so that the system after time $t$ is given by $\rho_t = \sum_{n=0}^{\infty}\exp \left (-\hbar\omega n/k_{\rm B}T \right )\ket{\Psi_n}\bra{\Psi_n} / Z$ with $\ket{\Psi_n}=\left(\ket{N}U_{\rm g}(t)+i\ket{N+1}U_{\rm e}(t)\right)\ket{n}/\sqrt{2}$. This involves
particle time-evolution operators associated with the ground and the excited state of the qubit,
\begin{equation}
\mathsf{U}_{\rm g}(t)=\mathsf{D}\left(\frac{V_{\rm ext}}{\hbar \omega}\right)\exp\left(-i\omega t a^\dagger a\right)\mathsf{D}\left(-\frac{V_{\rm ext}}{\hbar \omega}\right)
\end{equation}
and
\begin{eqnarray}
\mathsf{U}_{\rm e}(t) & = & \exp\left[-it\left(\frac{E_{\rm c}}{\hbar}-\frac{\kappa^2}{\omega} - \frac{2 \kappa V_{\rm ext}}{\hbar \omega} \right)\right]\mathsf{D}\left(\frac{\kappa}{\omega} +\frac{V_{\rm ext}}{\hbar \omega}\right) \notag \\
&& \times \exp\left(-i\omega t a^\dagger a\right)\mathsf{D}\left(-\frac{\kappa}{\omega} -\frac{V_{\rm ext}}{\hbar \omega}\right),
\end{eqnarray}
where  $\mathsf{D}(\alpha)=\exp\left(\alpha\mathsf{a}^\dagger-\alpha^*\mathsf{a}\right)$
is the displacement operator. 
The scheme with $\pi$-pulses at times $t_1$ and $t_2$ then results  in $\ket{\Psi_n}=\left(\ket{N}\mathsf{U}_++i\ket{N+1}\mathsf{U}_-\right)\ket{n}/\sqrt{2}$ at time $t_3$, where
\begin{subequations}
\begin{eqnarray}
\mathsf{U}_+ & = &\mathsf{U}_{\rm g}(t_3-t_2)\mathsf{U}_{\rm e}(t_2-t_1)\mathsf{U}_{\rm g}(t_1),\\
\mathsf{U_-} &= &\mathsf{U}_{\rm e}(t_3-t_2)\mathsf{U}_{\rm g}(t_2-t_1)\mathsf{U}_{\rm e}(t_1).
\end{eqnarray}
\end{subequations}
The charge occupation of the box after a final $\pi/2$-pulse is thus given by
\begin{equation}
\braket{\sigma_+\sigma_-}  =  \frac{1}{2}+\frac{1}{2Z}\sum_{n=0}^{\infty} \Re \left [\bra{n}\mathsf{U}_+^\dagger\mathsf{U}_-\ket{n}\right ] \exp \left ( -\frac{\hbar\omega n}{k_{\rm B}T}\right ).
\end{equation}
Noting that $\mathsf{U}_+^\dagger\mathsf{U}_-$ displaces the particle state in phase space and using
\begin{equation}
\begin{split}
&\frac{1}{Z} \sum_{n=0}^{\infty}\bra{n}\mathsf{D}(\alpha)\ket{n}\exp \left (-\frac{\hbar\omega n}{k_{\rm B}T} \right ) \\
&=\exp\left[-\coth \left ( \frac{\hbar \omega}{2 k_{\rm B} T} \right )\frac{|\alpha|^2}{2}\right],
\end{split}
\end{equation}
\tcb{one obtains
\begin{eqnarray}\label{measurementoutcome}
\braket{\mathsf{\sigma}^+\mathsf{\sigma}^-} &=&\frac{1}{2}+\frac{1}{2}\exp\left[-\frac{\kappa^2}{2\omega^2}\coth \left ( \frac{\hbar \omega}{2 k_{\rm B} T} \right )\vert d(t_1,t_2,t_3)\vert^2 \right]
\notag\\
&&\times\cos\left[\frac{\kappa}{\omega} \left (  \frac{\kappa}{\omega}+ \frac{2V_{\rm ext}}{\hbar \omega} \right ) {\rm Im}\,d(t_1,t_2,t_3) \right. \notag \\
& &\left. -\left(\frac{\kappa^2}{\omega} + \frac{2\kappa V_{\rm ext}}{\hbar \omega}-\frac{E_{\rm c}}{\hbar}\right) ( 2 t_1 - 2 t_2 + t_3) \right] ,
\end{eqnarray}
where $d(t_1,t_2,t_3)=2 e^{i\omega t_1}-2e^{i\omega t_2}+e^{i\omega t_3} - 1$.
Qubit dephasing in Fig.~\ref{figure3} is modeled by taking $n_g$  in (\ref{eq:hamtot}) to be a random number with Lorentzian distribution of width $\gamma_{\rm d}C_\Sigma\hbar/(2e)^2$. This adds the exponential factor $\exp(-\gamma_{\rm d} t_3)$ to the second term in (\ref{measurementoutcome}).

Equation \eqref{measurementoutcome} shows that the envelope of the qubit occupation assumes its maximum at $d(t_1,t_2,t_3) = 0$ and that the particle temperature determines the width of the peak. Operating the interference protocol at the point of maximal envelope (corresponding to a maximal overlap in the particle state of both superposition branches) can be achieved  with the symmetric choice $t_1 = \tau$, $t_2 = \tau + \Delta_\tau$ and $t_3 = 2 \tau + \Delta_\tau$, where 
\begin{equation}
\Delta_\tau = \frac{1}{\omega}\arctan\left[\frac{2 \sin(\omega \tau) [2 - \cos(\omega\tau)]}{[2 - \cos(\omega\tau)]^2 - \sin^2(\omega \tau)}\right]
\end{equation}
and $\tau<\pi/\omega$. Evaluating \eqref{measurementoutcome} for these times finally yields \eqref{measurementoutcome2}.
}

\nsubsection{Generation of nanoparticle entanglement}
\noindent
The time evolution involving two nanoparticles can be described by means of the respective particle operators 
\begin{subequations}
\begin{eqnarray}
\mathsf{U}_+^{(i)}\ket{n}&=&e^{i\phi_+^{(i)}}\mathsf{D}_i(\alpha_{i})\ket{n},\\
\mathsf{U}_-^{(i)}\ket{n}&=&e^{i\phi_-^{(i)}}\mathsf{D}_i(\beta_{i})\ket{n}.
\end{eqnarray}
\end{subequations}
where the ${\sf D}_i$ are phase-space displacement operators acting on nanoparticle $i$.

To entangle the particles one prepares the qubits in a Bell state and  performs the trapped interference scheme on both subsystems. Measuring both qubits before the wave packets overlap, e.g.\ at $\tau+\Delta_\tau<t_3<2\tau+\Delta_\tau$ projects
the two oscillators onto the outcome-conditioned state
\begin{equation} \label{condstate}
\rho'\propto\sum_{n,m=0}^\infty \exp \left [-\frac{\hbar\omega}{k_{\rm B}T}(n+m) \right ]\ket{\Psi_{nm}}\bra{\Psi_{nm}},
\end{equation}
where
\begin{equation}\label{entangledparticlestate}
\ket{\Psi_{nm}}=\left [\mathsf{D}_1(\alpha_1)\mathsf{D}_2(\beta_2)\pm e^{i\phi}\mathsf{D}_1(\beta_1)\mathsf{D}_2(\alpha_2)\right ]\ket{n}_1\otimes\ket{m}_2.
\end{equation}
The amplitudes $\alpha_i$, $\beta_i$ and the phase  $\phi=\phi_-^{(1)}+\phi_+^{(2)}-\phi_+^{(1)}-\phi_-^{(2)}$ depend on the pulse times, whereas the sign in (\ref{entangledparticlestate}) is fixed by the outcome of the qubit measurements.
The values of $\alpha_i$ and $\beta_i$ determine the amount of entanglement of the state \eqref{condstate}, as quantified by a suitable entanglement measure.

\nsubsection{Experimental parameters}
\noindent
For calculating the interference pattern in Fig.~\ \ref{figure3} we consider a cylindrically shaped silicon nanoparticle (diameter of $4.7\,{\rm nm}$, length of $42\,{\rm nm}$, homogeneously charged with $q = 200\,$e \cite{draine1987collisional}. 
We assume a value of  $\vert \blg{p} \vert= 200\,{\rm e\AA}$ for the dipole moment, based on previous studies reporting values of several 10\,e\AA\ for neutral particles of the same size \cite{shanbhag2006origin,shim1999permanent,yan2010effects}.
The Paul trap, with an endcap distance of  $2z_0=0.5\,{\rm mm}$ and geometry factor $k=0.4$  \cite{itano1995,goldwater2018levitated}, is driven by an AC voltage of $U_{\rm ac}=1\,{\rm kV}$ with frequency $\Omega=2\pi\times250\,{\rm MHz}$.

The empty Cooper-pair box has a capacitance of $C_\Sigma=4.4\,{\rm fF}$, yielding a charge energy of $2e^2/C_\Sigma\approx 72\,{\rm \mu eV}$. A relatively high occupation $N=10$ shifts the potential minimum by the distance $z_{\rm s}=1.17\,{\rm \mu m}$ from the trap center. The fast box oscillations then require a measurement  time resolution on the ${\rm ps}$-scale \cite{nakamura1999coherent}. The total duration of the experiment of $87\,{\rm ns}$ is on the expected coherence time scale of a charge qubit  \cite{houck2009life}.

An initial motional temperature of the particle of $T=1\,{\rm mK}$ is achievable via resistive cooling \cite{goldwater2018levitated,clark2010method,iftikhar2016primary} (and potentially by  electric feedback cooling \cite{tebbenjohanns2019cold,conangla2019optimal,goldwater2018levitated} or optical techniques \cite{delic2019motional}). Assuming a resistance of $R=100 \,{\rm M\Omega}$, the adiabatic cooling rate is $163\,{\rm Hz}$, corresponding to $1.54\times 10^5$ quanta per second in thermal equilibrium. Our estimate of the surface noise \cite{lakhmanskiy2019observation}  for the present system  with superconducting endcap electrodes yields a heating rate of $170\,\hbar\omega/$s, which does not noticeably raise the temperature of the particle on the time scale of the experiment.

\end{document}